\def\lapprox{{_<\atop{^\sim}}}

\def\cmmt{\rm {cm^{-2}}} 
 
\def\s-1{\rm {s^{-1}}}

\def\etal {et al.}

\def\kms {\hbox{${\rm km\,s}^{-1}$}}
\def\permpc {$\rm {Mpc^{-1}}$}

\def\pcsquare {$\rm {pc^{-2}}$}
\documentclass{aa} 
\usepackage{graphicx}

\begin{document}  
\title{Star-formation in the central kpc of the starburst/LINER galaxy NGC~1614}  
\author{E.~Olsson\inst{1}, S.~Aalto\inst{1}, M.~Thomasson\inst{1}, R.~Beswick\inst{2}} 
\offprints{S. Aalto} 
\institute{Chalmers University of Technology, Department of
Radio and Space Science, Onsala Space Observatory, SE-43992 Onsala, Sweden
\and University of Manchester, Jodrell Bank Centre for Astrophysics,
Oxford Road, Manchester, M13 9PL, UK}  
\date{Received / Accepted }  
\titlerunning{Star-formation in the central kpc of the starburst/LINER galaxy NGC~1614} 
\authorrunning{Olsson \etal}


\abstract
   {}
{The aim is to investigate the star-formation and LINER (Low Ionization Nuclear 
Emission Line Region) activity within the central kiloparsec of the galaxy 
NGC\,1614. In this paper the radio continuum morphology, which 
provides a tracer of both nuclear and star-formation activity, and the 
distribution and dynamics of the cold molecular and atomic gas feeding 
this activity, are studied. In particular, the nature of an 
R$\approx$300\,pc nuclear ring of star-formation and its relationship to 
the LINER activity in NGC\,1614 is addressed.}
{A high angular resolution, multi-wavelength study of the LINER 
galaxy NGC\,1614 has been performed. Deep observations of the CO 1-0 
spectral line were performed using the Owens Valley Radio Observatory 
(OVRO). These data have been complemented by extensive multi-frequency 
radio continuum and H{\sc i} absorption observations using the Very Large 
Array (VLA) and Multi-Element Radio Linked Interferometer Network (MERLIN).}
{Toward the center of NGC\,1614, we have detected a ring of radio 
continuum emission with a radius of 300\,pc. This ring is coincident with 
previous radio and Pa$\alpha$ observations. The dynamical mass of the ring based on H{\sc i}
absorption is $ 3.1 \times 10^9  \, \rm M_{\odot}$. The peak of the integrated CO 1-0 emission is shifted by
1$''$ to the north-west of the ring center. An upper limit to 
the molecular gas mass
in the ring region is $\sim1.7 \times 10^9  \, \rm M_{\odot}$. 
Inside the ring, there is a north to south elongated
1.4 GHz radio continuum feature, with a nuclear peak. This peak is
also seen in the 5 GHz radio continuum  and in the CO.}
{We suggest that the $R=300$ pc star forming ring represents the radius of 
a dynamical resonance - as an alternative to the scenario that the starburst is
propagating outwards from the center into a molecular ring. 
The ring-like appearance is probably part of a spiral
structure. Substantial amounts of molecular gas have passed the radius of
the ring and reached the nuclear region. The nuclear peak seen in 5 GHz radio continuum and CO
is likely related to previous star formation, where all molecular gas
was not consumed. The LINER-like optical spectrum observed in NGC~1614
may be due to nuclear starburst activity, and not to an Active Galactic Nucleus (AGN).
Although the presence of an AGN cannot be excluded.}


\maketitle

\section{Introduction}

NGC~1614 is a luminous infrared galaxy (LIRG) at a distance of 64 Mpc
(for $H_0$=75 \kms \permpc). The galaxy is barred and interacting (morphologically classified as
type  {SB(s)c~pec}) and its nuclear optical spectrum shows both starburst
and LINER (Low Ionization Nuclear 
Emission Line Region) activity. Although there is still no single consensus to
what is powering the emission in LINERs,  nuclear starburst or AGN (Active Galactic Nucleus)
activity are two major candidates (e.g. Ho 1999; Terashima
\etal\,
2000; Alonso-Herrero \etal\, 2000). 
Bar-driven
inflow of gas in galaxies  has been suggested to trigger nuclear
starbursts as well as feeding AGNs (e.g. Simkin, Su \& Schwarz 1980;
Scoville \etal~1985).  
However, the radial inflow of gas
along the bar may be slowed down at certain radii, often associated
with inner Lindblad resonances (ILR) (Combes 1988; Shlosman, Frank \&
Begelman 1989).

\par
Previous studies of NGC~1614 include e.g. an optical,
near-infrared, radio continuum and {\sc{hi}} study by Neff \etal~(1990).
They suggested that the spectacular structure with tidal tails
or plumes is the result of an earlier interaction with another galaxy of
comparatively modest mass and impact velocity. Alonso-Herrero \etal~(2001),
hereafter AH2001, have investigated  NGC~1614, in the optical and
near-infrared. They detected a starburst nucleus of about 45 pc in
diameter based on deep CO stellar absorption. The nucleus is surrounded by a
$\sim$600 pc diameter ring of current star formation, which is revealed
in Pa$\alpha$ line emission. Just outside the star forming ring, a
dust ring is indicated by its extinction shadow in $H-K$. Neither
Neff \etal~(1990) nor AH2001 find any indication of
an AGN. Scoville \etal~(1989) used the  Owens Valley Radio Observatory (OVRO),
which contained three elements at that time, to map NGC~1614 in CO
1--0 at a resolution of $4'' \times 6''$. They found an unresolved
CO-concentration with a mass of 
 $6 \times 10^9 $
M$_{\odot} $.

\begin{figure*} 
\centering 
\includegraphics[width=17cm]{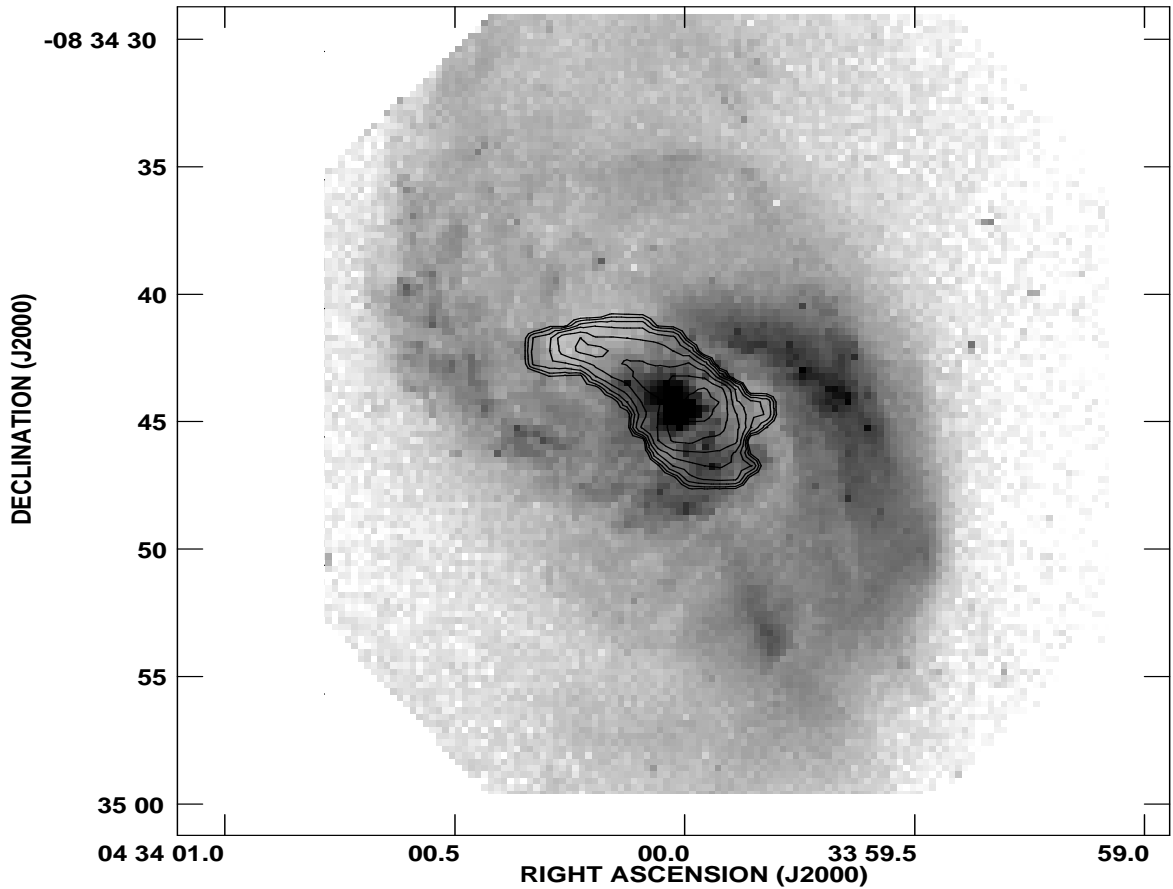} 
 \caption{Overlay of OVRO CO 1--0 integrated intensity contours over a
 F606W  WFPC2 {\it{HST}} image. The contours are in percent of the peak value of 34
 Jy\,km\,s$^{-1}$ per beam of $2.''75\times2.''40$. The levels are 8,
 11, 16, 23, 32, 45, 64 and 90\%. The greyscale of the optical image is
 logarithmic and arbitrary. The astrometrical alignment of the {\it{HST}} image
 with the CO map is within $1.''5$ (see Sect.~\ref{coresults}).}
\label{hst} 
\end{figure*}

We have used OVRO to study the distribution and dynamics of the
molecular gas at $2''$ resolution. We have
also studied the morphology of the 1.4 GHz radio continuum as well as
the distribution and dynamics of the neutral gas via H{\sc i} in
absorption at arcsecond resolution, using the Very Large Array (VLA). To study the
morphology of the nuclear radio continuum, we used the Multi Element
Radio Linked Interferometer Network
 (MERLIN) and obtained maps at higher (0.5 arcsecond)
 resolution, at 1.4 GHz and 5 GHz. 
The purpose of the studies was to 
provide information about the feeding and nature of the central activity giving rise
 to the LINER like spectrum in NGC~1614. In particular, we would like to
 address the 
question whether the LINER activity is due to an AGN or to a nuclear starburst.

\section{Observations and data reduction} 

\subsection{Owens Valley Radio Observatory} 
The OVRO mm interferometer was used to map NGC~1614
in CO 1--0 in the inner $60''$ ($R=9.3$ kpc). The array consists of six
10.4 m telescopes and was  used in the equatorial and high  resolution
configuration. These observations were carried
out in February 1996. 
The NRAO's {\sc aips} software package was used to
deconvolve the images. We used  data from both configurations
together to produce maps with high sensitivity and image fidelity, and data from the high resolution configuration only to produce maps
with high angular resolution. The resulting low and high resolution synthesized beams were $4.''44\times4.''10$ and $2.''55\times2.''40$.
The primary beam diameter was $60''$. 
The digital correlator was
centered at 113.4 GHz (4800 \kms) and was configured to cover 448 MHz (1200
\kms), with 8 MHz (22 \kms) resolution.
Typical system temperatures were 400~K.

\begin{figure*} 
\centering 
\includegraphics[width=18cm]{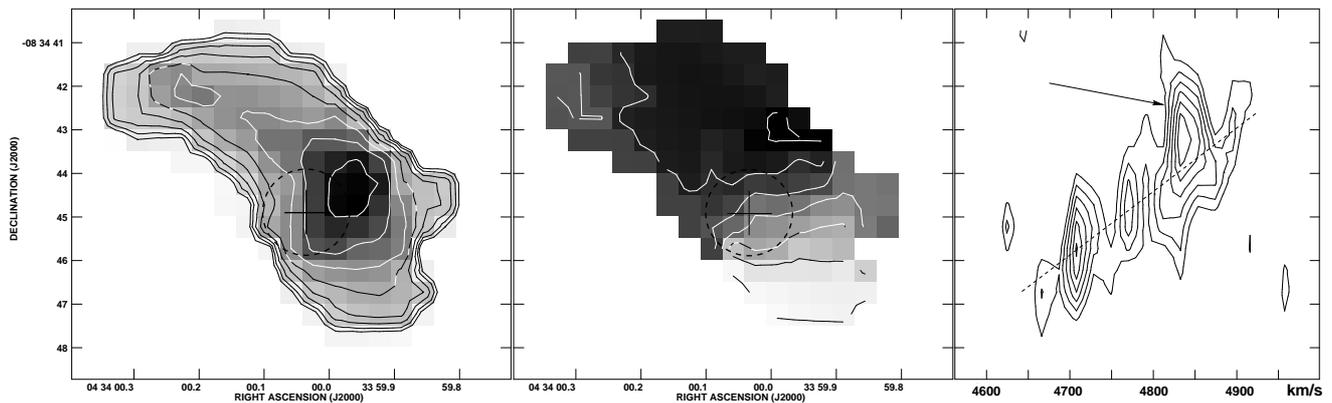} 
 \caption{OVRO CO 1--0 integrated intensity, velocity field  and major axis (north to
 south) position velocity
 diagram. The integrated intensity  \textit{(left panel)} contours are in percent of the peak value of 34
 Jy\,km\,s$^{-1}$ per beam of $2.''75\times2.''40$. The levels are 8,
 11, 16, 23, 32, 45, 64 and 90\%. The greyscale range
 is from 0 to 34 Jy\,km\,s$^{-1}$\,beam$^{-1}$. The velocity field
 \textit{(middle panel)} greyscale and contour range is from 4700 to 4850
 km\,s$^{-1}$, with 25 km\,s$^{-1}$ contour increments.
The position velocity diagram \textit{(right panel)}  is 
averaged over a $2.''5$ slit, centered on
the peak of the integrated
intensity (see Table 1). The dotted line represents the rotation curve in this region,
 and the arrow points at an additional kinematical component. The position velocity diagram contours are 40,
 50, 60, 70, 80 and 90\% of the peak value of 0.28 Jy per beam. The dotted ring marks the position of the 
starburst ring. Note that it is offset to the east from the CO peak by 1$''$.}
\label{comomsandpv} 
\end{figure*}

\subsection{Very Large Array}
\label{obsvla}
NGC 1614 was observed at 1.4 GHz with the VLA in the
A-configuration in April 2006. The observations were interspersed with
regular observations of the nearby phase calibrator 0423--013. The flux
calibrator was 3C\,286. The observations were centered on 1398 MHz,
which corresponds to a velocity of 4778  \kms.  A two-IF mode was used,
with a band width of 6.25 MHz per IF. The IFs were parallel in
frequency, which resulted in a total frequency
coverage of 6.25 MHz (1360 \kms). Each IF consisted of 32 channels with
a bandwidth of 96 kHz, which corresponds to a velocity resolution of
21.3 \kms. The data were  edited and calibrated  with {\sc{aips}} standard
procedures. The calibrated data were Fourier transformed to produce
spectral line data cubes with two different weightings. The line free
channels were combined to form continuum images, which were then
subtracted from the spectral line data. The continuum images and
the continuum subtracted spectral line data  were separately
deconvolved with the  {\sc{aips}} task {\sc{apcln}}, and were then
recombined to form the deconvolved spectral line data cubes. Two
different weighting schemes were applied to the data in order to obtain
maximum sensitivity and angular resolution. For maximum sensitivity,
natural weighting was applied, which resulted in a synthesized beam of
$2.''33\times1.''42$. For maximum angular resolution, a more uniform
weighting (robustness parameter $-2$) was applied. The resulting
synthesized beam was $1.''39\times1.''01$. Further analysis such as
presenting cleaned contoured continuum images, absorption spectra and
moment maps were done with  standard {\sc{aips}} tasks. 

\par

In order to provide complementary high resolution radio continuum imaging 
of the central region of NGC\,1614, data at 5 and 8.4\,GHz was  
obtained from the VLA archive and re-imaged. These short snapshot data 
were observed on 27th July 1999 in the VLA's highest resolution 
A-configuration. Each data-set was was calibrated using standard data 
reduction techniques, including phase referencing using a nearby 
calibrator, within the {\sc aips} packages. The flux density scale of 
these observations was calibrated with respect to 3C\,286 using the 
Baars \etal~(1977) scale.

\subsection{Multi Element Radio Linked Interferometer}
MERLIN was used in August 2005 to observe NGC~1614 at 1420 MHz. The
observations were interspersed with regular observations of the nearby
phase calibrator 0436--089. 3C\,286 was used as the primary calibrator
and 0552+398 as the secondary. They were both observed at the beginning
and end of the observing run. Dual bands of circular polarization were
recorded over a total bandwidth of 8 MHz, which was correlated into 64
channels width a bandwidth of 125 kHz each, which equals a velocity
resolution of 26 km\,s$^{-1}$. Initial editing and calibration of the
data was done at Jodrell Bank in September 2005 using the local MERLIN
{\sc{dprogs}} software. These data were read into {\sc{aips}} and further
calibration was done using the MERLIN pipeline, which included several
cycles of self calibration on the phase calibrator. Our target source
was not itself suitable for self calibration, following the constraints
given in the MERLIN handbook, so the phase corrections derived from the
phase calibrator were applied to our target source in the MERLIN
pipeline. The calibrated  {\it 
uv} data-set was Fourier transformed with no
deconvolution initially applied. The line free channels were combined to
produce continuum images which were used to subtract the continuum
contribution in the spectral line cubes. The continuum images and the
continuum subtracted spectral line data cubes were separately
deconvolved with the  {\sc{aips}} task {\sc{apcln}}, and were then
recombined to form the deconvolved  line data. The rather
low declination of this source resulted in an elongated synthesized beam
of $0.''50\times0.''17$. 

\par

In addition to the 1420\,MHz MERLIN data, MERLIN observations at 4994\,MHz 
were also obtained. These data consist of two full track 
observations of NGC\,1614, made on 24th April and 19th May 2000. In each 
case observations of NGC\,1614 were interspersed with scans of the phase 
reference source 0436--089, with additional observations of standard point 
source and flux density calibration sources at either end of the observing 
run. In total NGC\,1614 was observed for 12\,hr over the two observing 
dates. Both of these observations were independently calibrated and imaged 
using standard routines before the data were combined for final imaging 
of the target.

MERLIN has a shortest baseline spacing of $\sim$11\,km. As a consequence 
5\,GHz observations with MERLIN are insensitive to diffuse radio 
structures larger than $\sim$1\,arcsec. In order to restore these missing 
short spacing data, our MERLIN 5\,GHz data were combined within the {\it 
uv} plane with the calibrated VLA A-configuration snapshot data (see
Sect.~\ref{obsvla}), applying appropriate weightings to each data set
according to their 
sensitivity. The resulting combined data set was then deconvolved and 
imaged to provide maps both sensitive to the large diffuse radio emission 
and with an angular resolution intermediate between the individual MERLIN 
and VLA images.

\begin{table} 
\caption{\label{facts} Adopted properties of NGC~1614.}
\begin{tabular}{lr} 
Parameter & Value\\ 
\hline \\
CO-peak (J2000) & $ \alpha \,  04^{\rm h} 33^{\rm m}
59.96^{\rm s}, \delta \, -08^{\circ} 34' 44.7''  $  \\

Radio-peak (J2000) & $ \alpha \,  04^{\rm h} 34^{\rm m}
00.03^{\rm s}, \delta \, -08^{\circ} 34' 45.0''  $   \\

Optical peak (J2000)$^{\rm a} $ & $ \alpha \,  04^{\rm h} 34^{\rm m}
00.06^{\rm s}, \delta \, -08^{\circ} 34' 44.6''  $   \\

Morphological type & SB(s)c pec \\
Systemic velocity & 4778 \kms \\ 
Distance$^{\rm b} $ & 64 Mpc \\
Spatial scale & 1$''$ = 310 pc \\
Inclination$^{\rm c}$  & $51^{\circ}$\\
$L _{\rm IR} \,^{\rm d} $ & $3 \times 10^{11}$ L$_{\odot}$  \\
Adopted conversion factor & $2.3 \times 10^{20} $
$\cmmt $ (K \kms)$^{-1} $   \\
Adopted velocity convention &  Optical heliocentric \\

\hline \\
\end{tabular} \\  \\ 
a): Neff \etal, 1990    \\
b): For $H_0$=75 \kms \permpc \\ 
c): For the R=$1''$ P${\alpha}$ ring, Alonso-Herrero \etal, 2001 \\
d): Alonso-Herrero \etal, 2001  \\

\end{table}

\section{Results}

\subsection{The OVRO CO}
\label{coresults}
Figure~\ref{hst}
shows the high resolution integrated intensity contours overlayed on an
 F606W  WFPC2 (optical) {\it{HST}} (Hubble Space Telescope) image. The optical peak position has a
positional uncertainty of less than $1''$ (Neff \etal, 1990) and we
estimate that the astrometrical alignment between the optical and CO is
better than $1.''5$. The same CO integrated intensity map is
also shown in Fig.~\ref{comomsandpv} (left panel). The CO is centrally peaked, with an
extension to the northeast, with CO at lower integrated intensity. A
secondary peak is detected at $\sim3''$ (1 kpc) to the northeast. 

The maximum integrated intensity is 34 Jy \kms, in a beam of
$2.''75\times2.''40$. With a standard  CO to H$_2 $ conversion factor (see
table~\ref{facts}), this corresponds to a molecular mass of $1.3 \times 10^9 $
M$_{\odot} $ in the central beam. The projected surface density is 2600 M$_{\odot} $
\pcsquare\ in the central beam. The deprojected surface density is $\sim $
1600 M$_{\odot} $
\pcsquare\ for the adopted inclination of 51$^{\circ}$.

The high resolution velocity field map is shown in
Fig.~\ref{comomsandpv} (middle panel). The velocity  contours are
perpendicular to the main axis of the central CO structure, with
velocities ranging from 4675 to 4825 \kms.  On a linear scale, the
observed velocity gradient in the central $2''$ corresponds to $\sim$ 140 \kms
per kpc. In the extension to the northeast, the velocity contours are
more irregular, probably due to an additional kinematical component in
this region. In Fig.~\ref{hst}, this feature is spatially
consistent with a crossing dust lane.  

The position velocity diagram, along the north to south direction, is shown
in  Fig.~\ref{comomsandpv} (right panel). The position velocity diagram is
averaged over a $2.''5$ slit, centered on
the peak of the integrated
intensity. Three peaks are resolved. The two strongest peaks occur to
the north and south of the center. The third peak, which is weaker, occur toward the
center of the CO integrated intensity. 

The low resolution CO data do not show any additional features as
compared to the high resolution maps, but have higher sensitivity. The total
flux recovered is 80 Jy \kms, which corresponds to a molecular mass of  $3.0 \times 10^9 $
M$_{\odot} $. NGC~1614 was previously observed with OVRO (Scoville \etal,
1989) at lower resolution ($4'' \times 6''$), where a total of 103   Jy
\kms was detected.

The global star formation rate (SFR) is related to ${L_{\rm FIR}}$
(${L_{\rm FIR}} = 3 \times 10^{11}   \rm{L}_{\odot} $) and is given by
Kennicutt~(1998) as SFR=$ {L_{\rm FIR}} /  5.8 \times 10^{9} $ M$_{\odot}$ yr$^{-1} $. This
relation indicates an SFR of $\sim$ 52 M$_{\odot}$ yr$^{-1} $ in
NGC~1614. The star formation efficiency, SFE, is given by SFE=${L_{\rm
FIR}}/{M_{\rm H_2}}$ and is $\sim 100$ (in the unit $\rm{L}_{\odot} / \rm{M}_{\odot}  $ ). This is in good agreement with
the expected value of $\sim$ 100 for interacting galaxies, given by e.g. Young \etal~(1989);
Planesas \etal~(1997).

\subsection{The VLA and MERLIN Radio Continuum}
\label{radioresults}
\begin{figure} 
\centering 
\includegraphics[width=8cm]{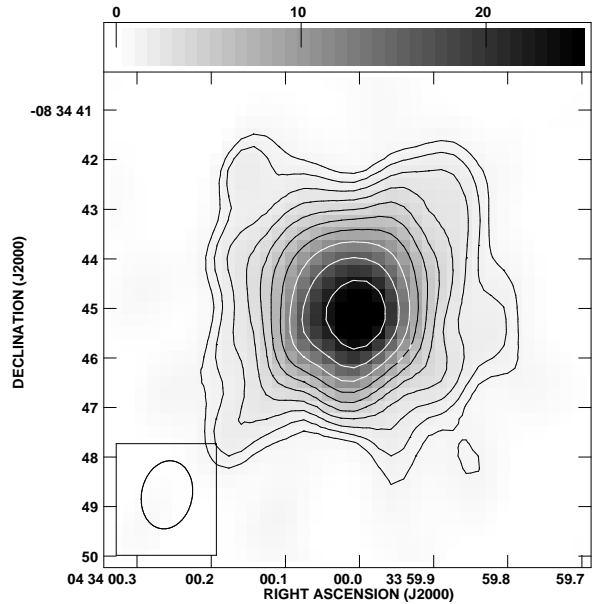} 
 \caption{Robustly weighted VLA radio continuum at 1.4 GHz. The
 synthesized beam is $1.''39\times1.''01$. The noise level is 0.32 mJy
 beam$^{-1}$. The contours start at 1 mJy and increase with a factor of
 $\sqrt{2}$ per level. The greyscale range is from 0 to 25 mJy
 beam$^{-1}$.}
\label{vlarobust} 
\end{figure}

Figure~\ref{vlarobust} shows the VLA
robustly weighted 1.4 GHz radio continuum map. The synthesized beam is
$1.''39\times1.''01$. The total flux density recovered is 130 mJy and
the noise level is 0.32 mJy beam$^{-1}$. 
The maximum flux in one beam is
29 mJy. In a naturally weighted 1.4 GHz radio continuum map (with a
synthesized beam of $2.''33\times1.''42$), we recover a total flux
density of 140 mJy.

\begin{figure*} 
\centering 
\includegraphics[width=17cm]{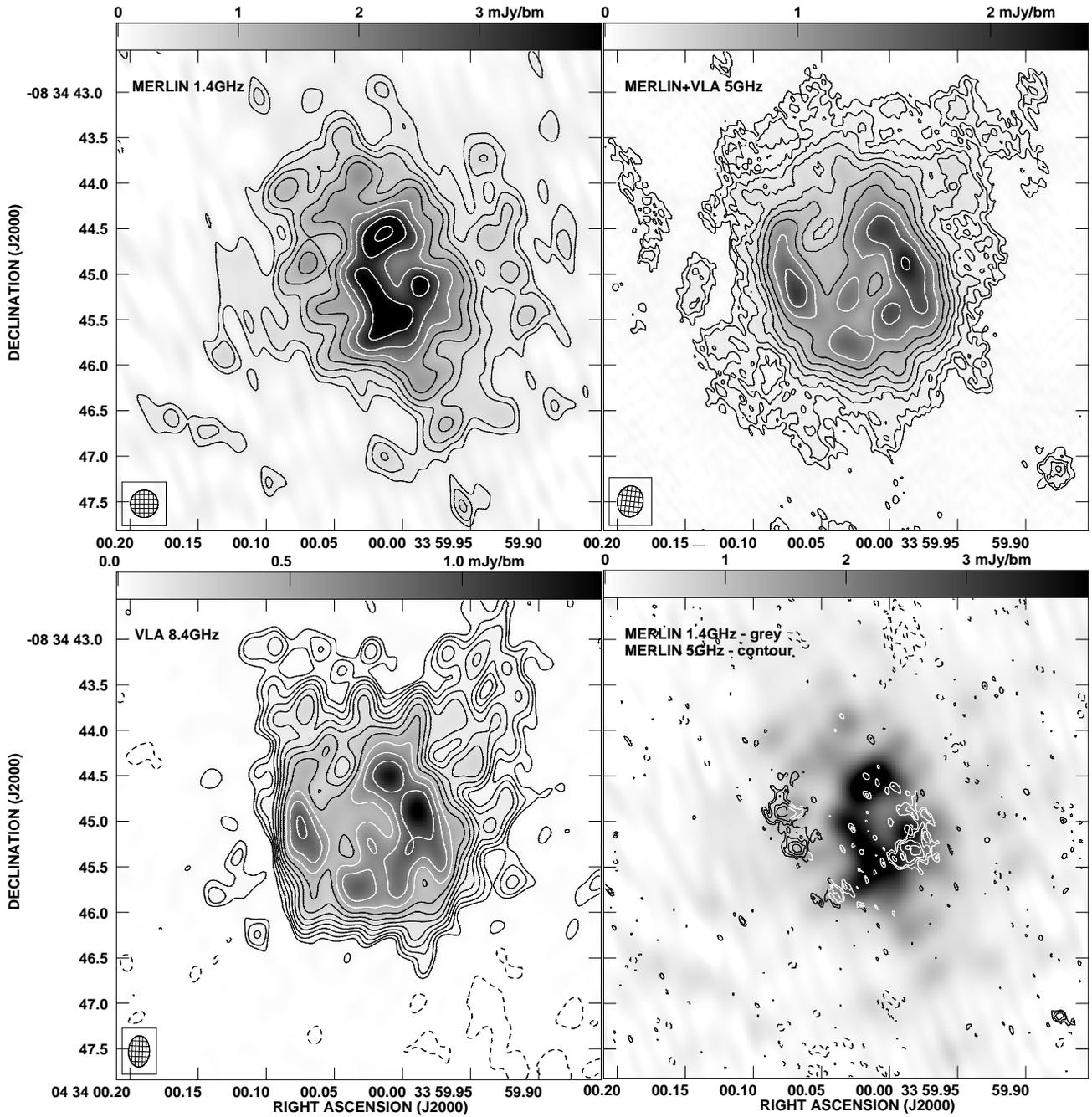} 
 \caption{\textit{Upper left panel:} MERLIN 1.4 GHz with a synthesized circular
 beam of $0.''3$. The greyscale is from 0 to 4 mJy
 beam$^{-1}$. The first contours are at $-0.32$ and 0.32 mJy
 beam$^{-1}$ and then increase with a factor of $\sqrt{2}$ per
 level. \textit{Upper right panel:} Combined MERLIN and VLA 5 GHz with
 a synthesized beam of $0.''34 \times 0.''28$. The greyscale is from 0 to 2.5 mJy
 beam$^{-1}$. The first contours are at $-0.1$ and 0.1 mJy
 beam$^{-1}$ and then increase with a factor of $\sqrt{2}$ per
 level. \textit{Lower left panel:} VLA 8.4
 GHz with
 a synthesized beam of $0.''34 \times 0.''23$. The greyscale is from 0 to 1.4 mJy
 beam$^{-1}$. The first contours are at $-0.029$ and 0.029 mJy
 beam$^{-1}$ and then increase with a factor of $\sqrt{2}$ per
 level. \textit{Lower right panel,
 greyscale:} MERLIN 1.4 GHz with a synthesized circular
 beam of $0.''3$. The greyscale is from 0 to 4 mJy
 beam$^{-1}$. \textit{Contours:}  MERLIN 5 GHz with a synthesized circular
 beam of $0.''2$. The first contours are at $-0.32$ and 0.32 mJy
 beam$^{-1}$ and then increase with a factor of $\sqrt{2}$ per
 level. }
\label{merlinandvla} 
\end{figure*}

\begin{figure} 
\centering 
\includegraphics[width=8cm]{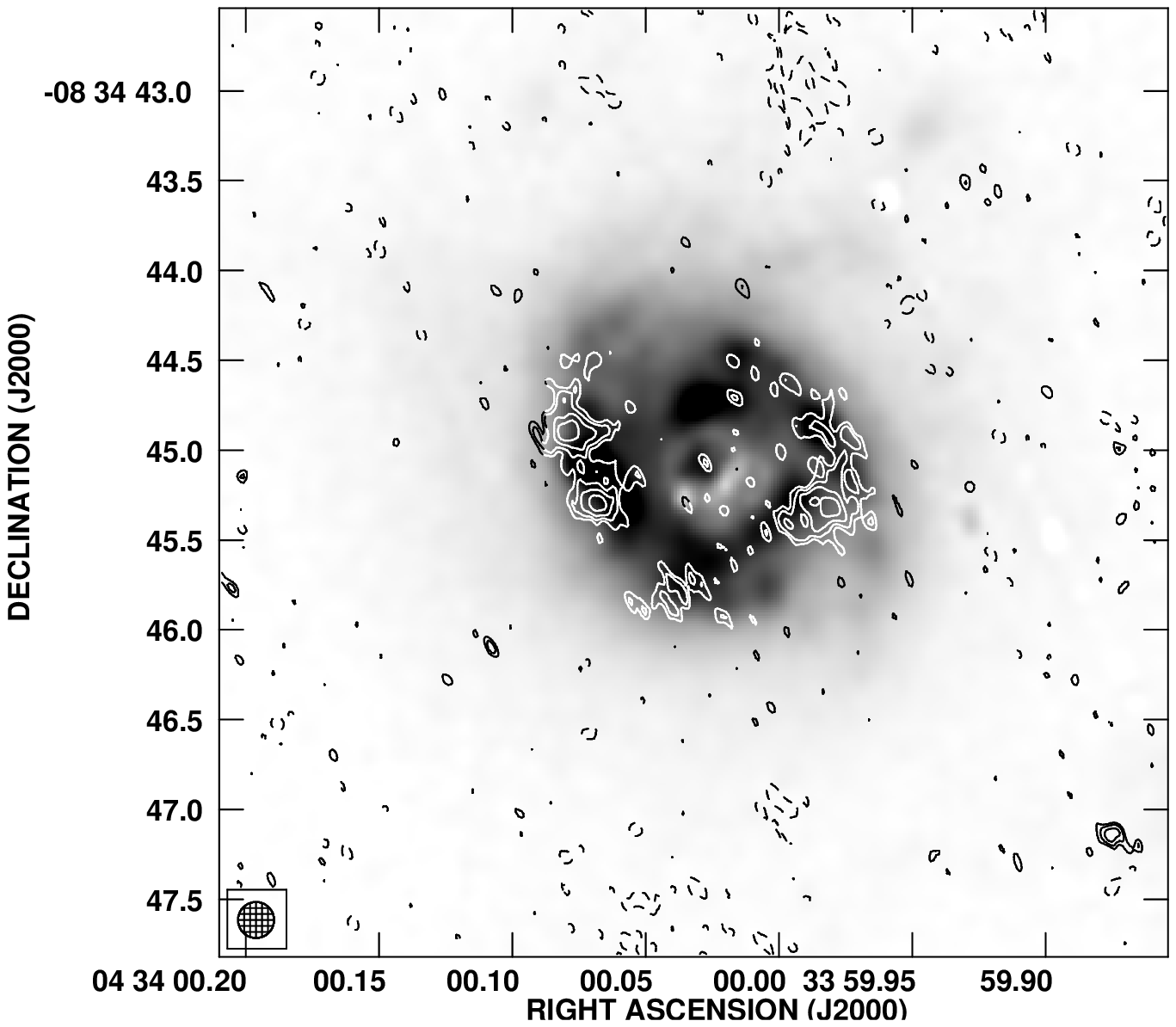} 
 \caption{Overlay of MERLIN 5 GHz in contours over a {\it{HST}} Pa$\alpha$ map in
 greyscale. The synthesized beam of the MERLIN data is circular with a
 diameter of $0.''2$. The dashed contour is at $-0.32$ mJy
 beam$^{-1}$, and the positive contours start at 0.32 mJy
 beam$^{-1}$ and has a contour increment of a factor of $\sqrt{2}$ per
 level. The greyscale is arbitrary.}
\label{merlinvlaandpaalpha} 
\end{figure}

Figure~\ref{merlinandvla} shows the MERLIN 1.4  GHz and 5 GHz radio
continuum of the central region of NGC~1614, as well as archival VLA 5
GHz 
and 8.4 GHz radio continuum of the same
region. Figure~\ref{merlinvlaandpaalpha} shows the MERLIN 5 GHz radio continuum
overlayed on a Pa$\alpha$ map\footnote{Based on observations made with the NASA/ESA
Hubble Space Telescope, obtained from the data archive at the Space Telescope Science
Institute. STScI is operated by the Association of Universities for Research in Astronomy,
Inc. under NASA contract NAS 5-26555} made with archival data from the Hubble
Space Telescope, {\it{HST}.}.

The Pa$\alpha$ (the greyscale in Fig.~\ref{merlinvlaandpaalpha}) traces
out a ring-like, patchy structure with a radius of $\sim 1''$ (310
pc). This has previously been reported by AH2001. Inside of this ring, the Pa$\alpha$ intensity is lower, except at
the center, where there is a  Pa$\alpha$ peak. We have spatially aligned
this peak with the position of the central peak of the  5 GHz MERLIN
radio continuum (the contours in Fig.~\ref{merlinvlaandpaalpha}).

The MERLIN 5 GHz map, the combined MERLIN and VLA 5 GHz map and the VLA
8.4 GHz map
all trace out the same ring-like structure as the Pa$\alpha$ (see
Fig.~\ref{merlinandvla}). We have chosen to refer to this
structure as the \textit{star forming ring}. The combined MERLIN and VLA
5 GHz map
(upper right panel, Fig.~\ref{merlinandvla}) 
coincides remarkably well in space with the  Pa$\alpha$, around the
whole circumference of the star forming ring. In the MERLIN map alone
(Fig.~\ref{merlinvlaandpaalpha} and lower right panel in Fig.~\ref{merlinandvla}), some of the emission
is resolved out, and mainly the eastern and western parts of the star
forming ring are detected. It is also detected in the VLA 8.4 GHz map (lower left panel,
Fig.~\ref{merlinandvla}). 

The MERLIN 1.4 GHz map (Fig.~\ref{merlinandvla}, upper left panel) reveals a north to south bar like structure, with several peaks
resolved. The two brightest peaks occur $1''$ (310 pc) from the center, 
on the star forming ring
toward the north and south respectively. The southern peak is spatially
extended to the north, toward the center. The third peak occurs toward the
west, in the Pa$\alpha$ ring.

There is an unresolved peak at the center of the star forming ring,
visible in most of our radio continuum maps. In particular, the peak is
isolated in the  MERLIN 5 GHz map, and tentatively
present in all maps.

\subsection{The VLA H{\sc i} Absorption}
\label{h1results}
\begin{figure} 
\centering 
\includegraphics[width=9cm]{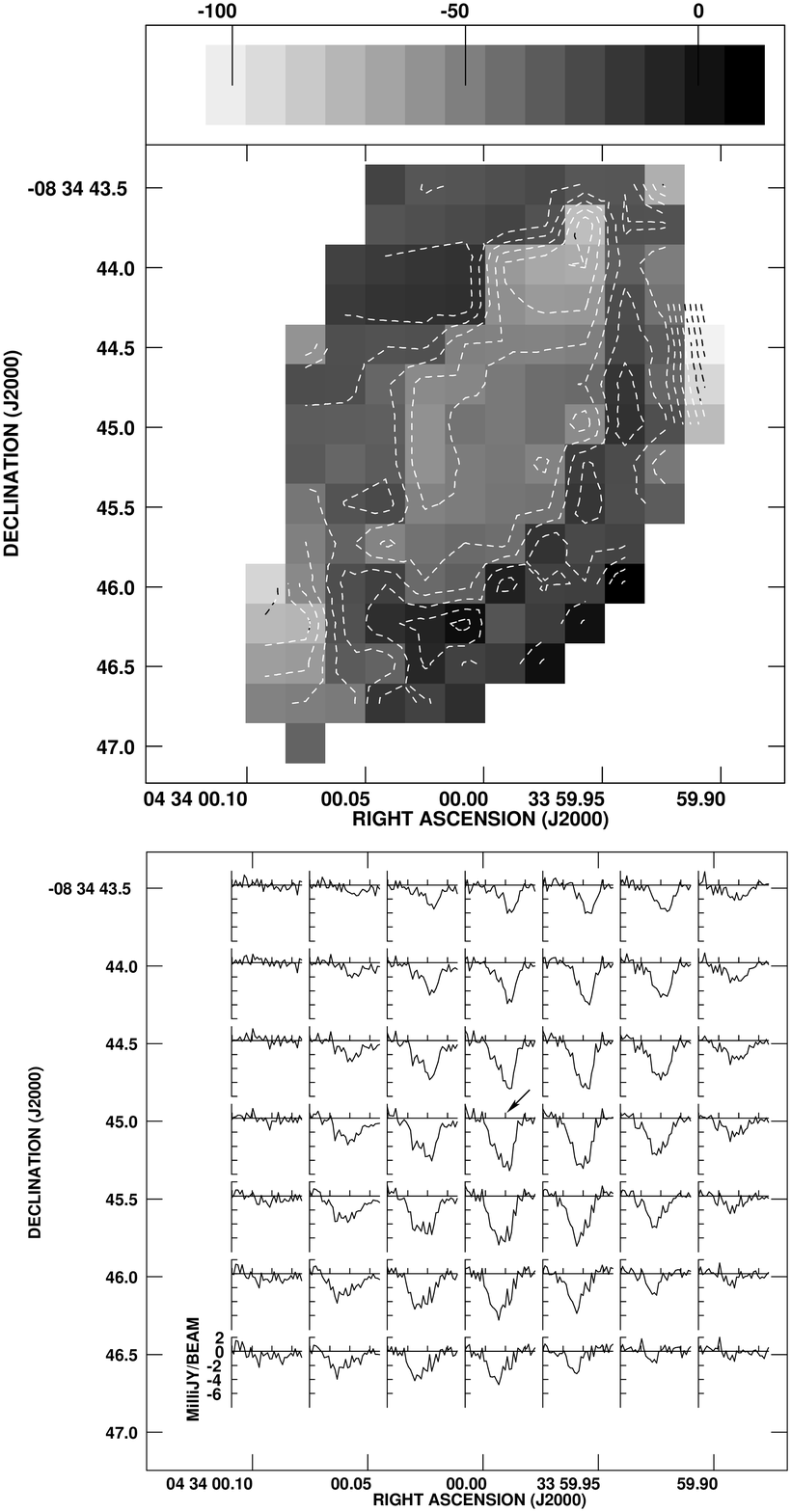} 
 \caption{\textit{The lower panel} shows spectra of the VLA naturally weighted
 continuum subtracted spectral line cube. The central spectra
 are at the radio continuum peak (see Table~\ref{facts}), and the
 separation between spectra is $0.''5$ in RA and DEC. The velocity scale
 is 200 \kms between ticks, and the tick marked with an arrow in the
 central spectrum is at 4800 \kms. The absorption scale is from $-8$ to 2
 mJy beam$^{-1}$, with increments between ticks of 2 mJy beam$^{-1}$. \textit{The upper panel} is the residual velocity field, after
 subtraction of a solid body rotation fitted only to the main (deeper
 than $-3$ mJy per beam) kinematical component. The greyscale and contour range
 is from $-110$ to 10 \kms and the contour increment is 10 \kms.} 
\label{spectraandresidual} 
\end{figure}

We have detected H{\sc i} in absorption in the central  $R \approx2''$ of
NGC~1614. Figure~\ref{spectraandresidual} (lower panel) shows spectra from the
VLA continuum subtracted spectral line data cube. 
H{\sc i} spectra from the MERLIN data (not shown) are consistent in width and 
optical depth with the  VLA spectra, but are not used for the analysis or
moment maps due to low signal to noise ratio.

The separation between
individual spectra in Fig.~\ref{spectraandresidual} is $0.''5$ in both RA and dec.  The absolute depth of the
absorption does not only depend on the column density of foreground
H{\sc i}, but also on the flux density of the background radio
continuum. Hence the absorption is deeper in the central spectra, and no
absorption is detected outside a radius of  $1.''5$ in the east to west
direction. To the north and south (not shown in this map) there are
tentative detections out to a radius of $2''$.

\begin{figure*} 
\centering 
\includegraphics[width=18cm]{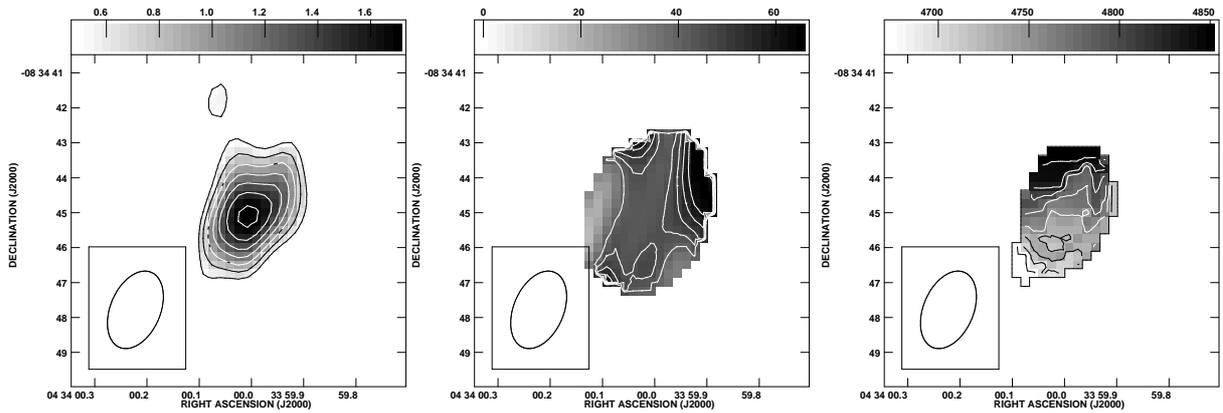} 
 \caption{\textit{The left panel} shows the integrated VLA H{\sc i} absorption,
 derived from the continuum subtracted spectral line cube. The greyscale
 and contour range are from $-500$ to $-1700$ mJy bm$^{-1}  \rm{km} \,
 \rm{s}^{-1}$. The contour increment is 200  mJy bm$^{-1}  \rm{km} \,
 \rm{s}^{-1}$. The synthesized beam is $2.''33\times1.''42$ (in all
 three panels). \textit{The middle panel} shows the integrated optical depth
 ($\int \tau \rm{dV}$). The greyscale range is from 0 to 65000 (in the
 unit \kms). The
 contour range is from 40000 to 60000 with increments of 5000. Note that
 the errors in this map increase in areas where the radio continuum is
 weak, i.e. toward the edges of the map. To minimize this effect, this
 map has been blanked where the radio continuum is less than 10 mJy
 bm$^{-1}$. \textit{The right panel} shows the velocity field, derived from the
 continuum subtracted spectral line cube. The greyscale range and
 contours are from 4675 to 4850 \kms with 25  \kms contour increments. }
 \label{himapsnatural} 
\end{figure*}

The moment maps of the H{\sc i} data
are shown in Fig.~\ref{himapsnatural}. The left panel is the integrated
absorption. Its peak occurs close to the radio continuum peak ( but is not a true measure
of the H{\sc i} distribution). 
The middle panel shows the
integrated opacity ($\int \tau \rm{dV}$). This map has been blanked
outside of the region where the radio continuum flux density is less
than 10 mJy beam$^{-1}$, since the noise in this map increases
dramatically where there is little or no radio continuum. The integrated
opacity is a measure of the foreground absorbing H{\sc i}, and there is
a clear east to west gradient in this map. 

The column density of absorbing atomic hydrogen, {$N\rm{_H}$}, has been
calculated at several positions using the equation

\smallskip

$N{\rm{_H}} = 1.823 \times 10^{18}T{\rm{_{spin}}} \int \tau {\rm{d}}V \,  \rm{atoms\, cm^{-2}}$

\smallskip
\noindent
where $T\rm{_{spin}}$ is the spin (excitation) temperature, and 
$\int \tau \rm{dV}$ is the integrated opacity (from
Fig.~\ref{himapsnatural}) in \kms. $T\rm{_{spin}}$ can be reasonably assumed to be 100
K (Maloney, Hollenbach \& Thielens, 1996), although it should be noted
that the value of $T\rm{_{spin}}$ is dependent on the physical
conditions of the gas.  

The maximum column density occur to the west and northwest of the
radio continuum peak, where the peak opacity is $\sim 0.25$ and the
velocity width is $\sim$ 200 \kms. These values result in a column
density approaching $10^{22} \, \rm{atoms\, cm^{-2}}$. East of the radio
continuum peak, the column density is approximately 50\% lower.

The right panel of Fig.~\ref{himapsnatural} shows the velocity
field. There is a clear north to south velocity gradient, where the southern end is approaching and the northern end
receeding, similar to what is seen in CO.

\subsection{Dynamical results with \sc{gal}}
\label{galresults}

We used the velocity of the OVRO CO data to fit a
solid body rotation curve to the inner $R= 5''$ with the {\sc{aips}}
task {\sc{gal}}. For solid body rotation, the inclination and rotational
velocity can not both be fitted with {\sc{gal}}, since they are directly
dependent on each other. We therefore kept the inclination fixed at
$i=51^{\circ}$, which is the inclination of the star forming ring (AH2001).
The dynamical center was kept
fixed at the CO-peak (see Table~\ref{facts}). The fit gave a
rotation of 40 \kms\, per arcsecond (130 \kms\, per kpc), which we used to
calculate a dynamical mass within a radius of $4''$ (1240 pc). We used
the Keplerian relation 

\smallskip

$M{_{\rm dyn}  = 2.3 \times 10^8  ({V_{\rm rot} \over100})^2
({R \over 100}) \, \rm M_{\odot}} $ 

\smallskip
\noindent
where $V_{\rm rot} $ is in \kms and $R$ in pc. This resulted in a
dynamical mass of $ 7.3 \times 10^9  \, \rm M_{\odot}$, which has to be
taken as an estimate since we can not rule out the presence of
non-circular orbits. The ratio of dynamical mass over molecular mass
detected in
the integrated intensity map is
$\sim 2.5$.

The residual velocity field, after the fitted rotation curve was
subtracted, was also calculated with {\sc{gal}}. The residual velocities range from $-40$ \kms to
50\kms, and the  zero point in the residual field occur toward the dynamical
center.

The velocity field derived from  the naturally weighted VLA continuum
 subtracted spectral line data cube is shown in Fig.~\ref{himapsnatural}
 (right panel). The main north to south rotation is clearly seen, but
 also a deviation in the velocity field in the northern half of the
 galaxy. As seen in Fig.~\ref{spectraandresidual} (lower panel), there are broad absorption
 line wings in several spectra.  In order to isolate the main rotation
 from additional kinematical components, we used the {\sc{aips}} task
 {\sc{gal}} to fit a solid body rotation curve to only the deepest, narrow
 absorption components. We used a cut off level of $-3$ mJy per beam to
 derive a main velocity field. We fitted a
 rotation curve, which was then subtracted from the
 original velocity field (Fig.~\ref{himapsnatural}, right panel). The
 fitted solid body rotation was $\sim$ 210 \kms\, per kpc, and the
 residual of the velocity field is shown in
 Fig.~\ref{spectraandresidual} (upper panel). The observed residuals are all blueshifted,
 with $\sim$ 100 \kms over most of the fitted area
 (the central $R=2''$).




\section{Discussion}

\subsection{Large scale molecular distribution and dynamics}
\label{codiscussion}
The integrated intensity and velocity field of the OVRO CO data is shown
in Fig.~\ref{comomsandpv} (left and middle panel). To the northeast, there is a CO extension, and
the gradient in the velocity field  falls, as compared to the gradient
in the center.  We believe that
this extension is caused by a separate, foreground gas component,
associated with a dust lane, which can be seen in Fig.~\ref{hst} (the
lighter region northeast of the nucleus). This component is likely not to be in the same 
plane as the spiral arms visible in the same figure.  For further discussion of this
feature see Sect.~\ref{dust_lane}.

 The position velocity diagram of the same data (Fig.~\ref{comomsandpv},
 right panel)
also shows an additional, blueshifted component which is associated with the
foreground dust lane. The dotted line in the position velocity diagram
was interpolated between the southern and central peaks (not contaminated by
the dust lane) and then extrapolated to the north. The additional
component is marked in the figure. We conclude that the bulk of the CO
is associated with a symmetric, slightly elongated bar like structure,
and that there is a second CO component associated with a foreground
dust lane to the northeast. The CO-contours in Fig.~\ref{hst} also shows
 an extension to the west of the main CO peak. 
This extension is consistent with also being 
associated with the same foreground dust lane, which
 crosses the galaxy from the northeast to the southwest. 

Note that the CO peak is offset from the radio continuum peak (and also
the starburst ring) by 1$''$ to the north-west (Fig.~\ref{comomsandpv}). This is consistent
both with the east-west extinction seen by AH2001 and the east-west gradient in opacity
in atomic hydrogen in absorption found in this paper in Sect.~\ref{h1results}. This suggests 
that the extinction shadow found by AH2001 is not a part of a complete ring but actually represents
a true lopsidedness in gas and dust around the center.

\subsection{The atomic gas -- column densities and dynamics}
\label{h1discussion}
\begin{figure} 
\centering 
\includegraphics[width=8cm]{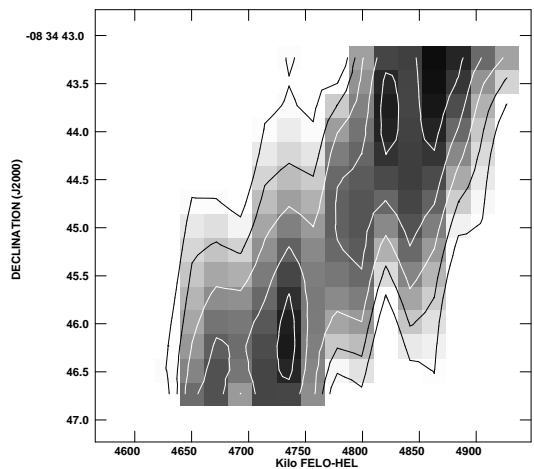} 
 \caption{Major axis (north to south) position velocity diagram of the
 VLA H{\sc{i}}
 opacity, $\tau$. The position velocity diagram is averaged over a
 $1.''75$ slit, centered on the radio continuum peak (see Table 1). The greyscale range is from 0.075 to 0.2. The contour
 range is from 0.075 to 0.175 with increments of 0.025.} 

\label{h1pvtau} 
\end{figure}

As described in  Sect.~\ref{h1results} and seen in the integrated
opacity map (Fig.~\ref{himapsnatural}, middle panel), the maximum column density of absorbing
H{\sc i} does not occur toward the radio continuum peak. To the
contrary, the highest opacity occur at the edges of the integrated
opacity map. This effect is better seen in the major axis position
velocity diagram of the opacity (Fig.~\ref{h1pvtau}), which clearly shows
two separate opacity peaks, to the north and south respectively. The
position velocity diagram is blanked where the background VLA radio
continuum flux density is less than 10 mJy beam$^{-1}$, which spatially
corresponds to $\sim$ 500 pc to the north and south of the radio
continuum peak. The peak opacities occur close to the edges of the
position velocity diagram, i.e. at  $R\approx$ 400 pc. This is 
consistent with, or slightly larger than, the radius of the star forming
ring described in Sect.~\ref{radioresults}, and we conclude that
most of the absorbing H{\sc i}  is associated with the outer parts of
this ring. The northern and southern peaks are also seen in the integrated opacity map (Fig.~\ref{himapsnatural}, middle
panel), which also shows a clear east to west opacity gradient, with a
peak to the northwest. This peak is spatially consistent with the
western end of the dust
lane (the light region northwest of the CO-peak in Fig.~\ref{hst}) and
is consistent with the extension seen in the CO-contours in this region.

In  Sect.~\ref{galresults}, we isolated the main H{\sc i} 
rotation, and showed that the  residual H{\sc i} velocity field is all
 blueshifted (Fig.~\ref{spectraandresidual}), 
with velocities of $\sim$ 100 \kms over most of the fitted area
 (the central $R=2''$).
 We interpret this as a large outflow, or super
wind, associated with the past and current star formation. Superwinds
 are found in almost all infrared-luminous galaxies (Rupke \etal\
 2005). For an inclination of $i=51^{\circ}$, our observations
 correspond to a deprojected wind velocity of  $\sim$ 160 \kms. This is
 in good agreement with the average outflow velocity of the ionized gas
 in LIRGs of 170 \kms (Lehnert \& Heckman, 1996).

\subsection{The central radio continuum -- a nuclear ring or spiral}

In  Sect.~\ref{radioresults}, we described a star forming ring with a
radius of $1''$ (310 pc), seen in the MERLIN and VLA radio continuum and
in the Pa$\alpha$ archival data. 
(Fig.~\ref{merlinandvla} \&~\ref{merlinvlaandpaalpha}).
It has been detected previously in the Pa$\alpha$, by AH2001, and 
with the VLA at lower resolution by Hummel \etal~(1987) and Neff \etal~(1990). 

\begin{figure} 
\centering 
\includegraphics[width=8cm]{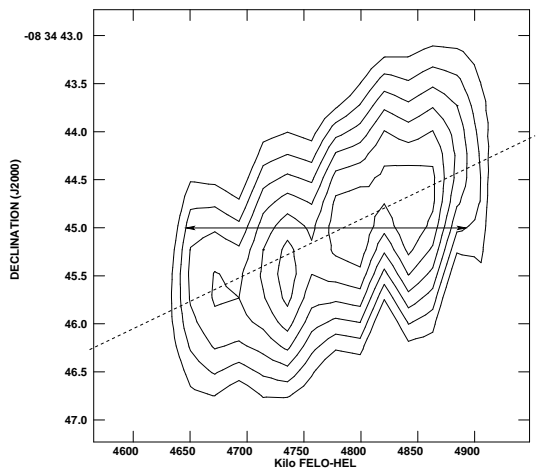} 
 \caption{Major axis (north to south) position velocity diagram of the
 VLA H{\sc{i}}
 absorption. The position velocity diagram is derived from the continuum
 subtracted spectral line data cube and is averaged over a
 $1.''75$ slit, centered on the radio continuum peak (see Table 1). The
  contours are at 40,
 50, 60, 70, 80 and 90\% of the peak absorption depth of $-6.4$ mJy per
 beam. The dotted line represents the rotation curve in this region. The
 arrow represents the central velocity width at 50\%  of the peak absorption depth.
} 
\label{h1pv} 
\end{figure}

The molecular mass in the central OVRO CO beam ($2.''75\times2.''40$) is $1.3 \times 10^9 $
M$_{\odot} $, using the conversion factor in Table~\ref{facts}. This mass can be used as an upper limit to the molecular
mass associated with the star forming ring. This is a conservative
upper limit since most of the CO peak appears not associated with the ring itself, but resides
1$''$ to the north-west of it. 
To make a mass budget for the ring, we have estimated the dynamical mass in the central
$R=1''$. The
highest possible spatial resolution was needed, and we used the VLA
H{\sc{i}} absorption data (with a synthesized beam of $2.''33\times1.''42$). 
Three peaks are resolved in the position velocity diagram (Fig.~\ref{h1pv}), and we estimated
 a projected
 velocity
 gradient in the ring of $\sim$170  \kms arcsec$^{-1}$, based on the slope of the dotted
 line in Fig.~\ref{h1pv}. A more conservative estimate would be to use
 the central line width at 50\% of the maximum absorption in the position
 velocity diagram (the arrow in Fig.~\ref{h1pv}), which is 240 \kms, and
 would give a projected
 velocity
 gradient in the ring of $\sim$120  \kms arcsec$^{-1}$  . For these two
 velocity gradients and an inclination of 51$^{\circ}$, using the Keplerian relation.

\smallskip
\noindent

$M{_{\rm dyn}  = 2.3 \times 10^8  ({V_{\rm rot} \over100})^2
({R \over 100}) \, \rm M_{\odot}} $ 

\smallskip
\noindent
where $V_{\rm rot} $ is in \kms and $R$ in pc, we derive a dynamical mass
range of  $ 1.7 \times 10^9  \, \rm M_{\odot}$ (conservative) to   $ 3.5
\times 10^9  \, \rm M_{\odot}$ in the central $R=1''$.

This is a factor of $\sim 1.5 - 2.5$
higher than the value calculated by AH2001. They reported
that the molecular gas mass, as calculated with a standard CO to H$_2$
conversion factor, would clearly exceed the dynamical mass of the star
forming ring. The  higher angular resolution
of our OVRO data, as compared to previous observations by Scoville \etal~(1989), has better isolated the CO in the star forming ring, and
provides a molecular gas mass 
estimate of $1.3 \times 10^9  \, \rm M_{\odot}$. This, and our
higher dynamical mass, allows a mass budget including all of the molecular gas detected in
the central OVRO beam.

Note that the CO peak is offset from the radio continuum peak by about 1$''$ suggesting
that most of the molecular mass in the CO peak is not associated with the star forming ring.
This further enforces the conclusion that there is no direct conflict between the estimated molecular
mass and the dynamical mass of the ring.

\subsection{Is there a bar in NGC~1614?}

The origin of the star forming ring can either be a nuclear starburst,
that has consumed most of the gas in the inner  $R=1''$, and is
progressing outwards, or it can have dynamical origin, possibly 
associated with an inner Lindblad resonance (ILR), where the gas would
pile up (Combes 1988a; Shlosman \etal\ 1989). NGC~1614 has previously been described as a textbook example of a
propagating starburst that started in the nucleus of a late-type, large
spiral galaxy, has grown outward to a radius of $\sim 300$ pc, and is
potentially still growing into a circumnuclear ring of molecular
material just outside this radius (AH2001).   

We suggest that, instead of being the result of a starburst spreading outward from
the center, the star forming ring represents the radius of a
dynamical resonance. The ring-like appearance may be part of a spiral structure,
extending from the nuclear region, inside of the ring, out to the spiral
arms visible in the F606W  WFPC2 {\it{HST}} image (Fig.~\ref{hst}). 

\begin{figure} 
\centering 
\includegraphics[width=8cm]{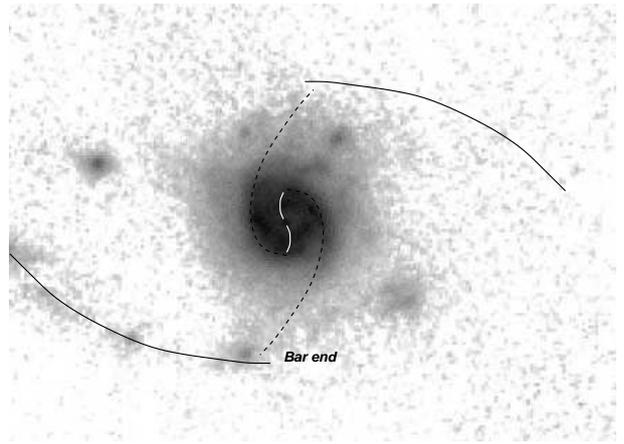} 
 \caption{Overlay of the main dynamical features (represented with arcs)
 over a Pa$\alpha$ map (same as in Fig~5) in greyscale. The solid black arcs represent the
 main spiral arms. The dashed arcs represent the leading edges of the
 main bar. The white arcs represent the putative inner decoupled bar. }

\label{model} 
\end{figure}

This scenario is shown in
Fig.~\ref{model}, where the greyscale is the Pa$\alpha$ emission and the
different dynamical features are marked with arcs. The outer black arcs represent the main
spiral arms, which are clearly visible in the  F606W  WFPC2 {\it{HST}} image and Pa$\alpha$
images (Fig.~\ref{merlinvlaandpaalpha} \&~\ref{model}). The dashed arcs represent the continuation of the main spiral
arms toward the center, probably associated with the leading edge of a
bar. We believe that the gas flows along the leading edge of the bar,
and piles up at the radius of the star forming ring. 

Inside the ring, there may be a dynamically decoupled bar, in the north to south
direction, which connects to the nucleus (the white arcs in
Fig.~\ref{model}). 
The gas mass fraction in the central $R=1''$ is well over 10\% which is the required
lower limit (Hunt \etal, 2008) for an
inner bar to decouple from the outer bar. The north-south feature in the
1.4 GHz MERLIN map (Fig.~\ref{merlinandvla}) may represent such an inner, decoupled bar.

The ring formed at the resonance acts as a barrier preventing further mass inflow, but Shlosman solved this by forming 
inner bars that may connect with the outer bar, i.e. nested bars  (e.g. Shlosman 2003).
A problem in our interpretation may then be that the nested bar forms after the ring formation and thus the
activity in the center should therefore be younger than the activity in the ring. However, there are at least two arguments
that counter this. Gas may flow to the nucleus before the ring is fully formed where it can feed an AGN or a starburst.
The nuclear activity may not be triggered immediately but is dependent on gas build-up times and dynamics. Furthermore, the
star formation in the ring itself may well be episodic and the ring could go through several bursts. 

Sarzi \etal\ (2007) discuss the formation of rings in systems both with and without strong kpc-scale bars. 
The morphology of the resonant nuclear rings in their sample is almost identical to the ring of NGC~1614 complete with a nuclear AGN and/or starburst. Sarzi \etal\ discuss how the formation of such rings may, instead of being caused by strong bars, originate due to 
a weak oval distortion or the tidal effects of a companion (e.g. Combes 1988b, Buta \etal\ 1995). Knapen \etal\ (2004) find
a nuclear ring in NGC~278 triggered by the non-axisymmetric potential induced by a recent minor merger.

Finding dynamically induced starbursts in interacting galaxies is not
unusual (e.g. Jogee \etal\, 2002;  Keto, Ho \& Lo 2005; ) where rings and inner spirals/bars are
associated with the locations of resonances. A prominent and well
studied example is the nearby starburst galaxy
M~82 which exhibits  a molecular ring with a radius of 200 pc as  seen in CO
1--0 (Walter, Weiss \& Scoville 2002). This ring has also been studied
in the highly excited CO 6--5 gas (Seaquist, Lee \& Moriarty-Schieven
2006) where it appears to be dynamically associated with the
transfer of gas from $x_1$ orbits in the large scale bar to perpendicular
inner $x_2$ orbits.

\subsubsection{Evidence for a barred potential}

Work by Chapelon \etal\ (1999) (C99) adresses the issues of
starbursts in barred spiral galaxies.  C99 claim that NGC~1614 is a barred galaxy with a bar PA of 37$^{\circ}$,
a bar length a of 5$''$ and a deprojected
a/b ratio of 0.23. C99 based their conclusions on optical CCD images and it is interesting to note that this PA is similar to
that for the crossing dust lane and CO distribution.

Limits on the parameters of a possible bar in NGC~1614 can be put on the onset of
the prominent two-armed spiral structure. Defining where the arms actually start is
not straight forward - but an estimate from the HST image suggests 7$''$ as a
projected limit. 
The prominent dust lane crossing just north of the central regions of NGC~1614 makes it difficult
to judge the stellar structure and backbone of the galaxy in the visible band. We therefore investigated
a 2.2 micron K-band image with 0.3 arcsecond resolution (from Calzetti 1997). 
The K-band image reveals that between a radius of 1.1 and 0.4 kpc there is an apparent oval
distortion of a/b axis ratio of 0.66 (and PA of 0 degrees) dropping to unity at radius 0.4 kpc. This is roughly where the 
nuclear ring commences. There is thus either a weak north-south bar with projected length 2.2 kpc,
or there is a warp in the inner kpc. Warps can occur in interacting systems. One example is the
warped gas of NGC~3718 (Krips \etal, 2005). One difference here is then that the dust lane of NGC~1614 would be 
at an angle to the warp - and not part of the warped structure - as it is in NGC~3718.
If there is a warp in NGC~1614 it would have to be gentle (because of the high axis ratio) but still
happen rather quickly (from 1.1 to 0.4 kpc). Thus, instead of a bar with PA 37$^{\circ}$ suggested by C99 we
propose a north-south bar based on the K-band image.

The CO velocity field suggests an S-shaped distortion in the iso-velocity contours. This is typical of
barred galaxies - but may also be the results of a warp (or ouflows or other non-circular motions).
The CO morphology is strongly affected by the crossing dust lane but is consistent with a north-south structure.
It is important to note that the CO peak and distribution is offset from that of the starburst ring.
The CO peak is located 1$''$ north-west of the radio continuum peak (and center of the radio continuum ring).
This is to some degree caused by the crossing dust structure - but not entirely. There is also a compact
feature at $v$=4870 \kms that helps to shift the integrated CO peak to the north-west. This could be the result of orbit crowding,
but higher resolution observations are necessary to determine the inner CO dynamics of NGC~1614.
It is furthermore interesting to note that in both the CO and H{\sc i} velocity field the velocity contours make
a twist at the location of the starburst ring resulting in a local PA of 30-40 degrees (see Fig.~\ref{comomsandpv}). 
Such a PA is consistent with the shape of the P$\alpha$ ring morphology.
Again, high resolution CO data are necessary to determine both the CO morphology and dynamics across
the ring and in the surrounding structure.

\subsection{Nuclear radio continuum morphology and star formation}
\label{nuclearradio}
Figure~\ref{merlinandvla} (lower right panel) shows the 1.4 GHz MERLIN
in greyscale and the 5 GHz MERLIN in contours.  The star forming ring is
clearly seen in the 5 GHz data, while the nuclear region is better shown
in the 1.4 GHz data, which has one of its peaks close to the center of
the star forming ring.

The strongest  1.4 GHz peak (the northern one) is 5.5 mJy in a circular
beam of $0.''3$. This can be converted to a source temperature of
53 000 K. This is an upper limit for the central peak, at this
resolution, and is consistent with a scenario of a cluster of SNRs in
the ring. However, an AGN can not be ruled out based on source
temperature at this resolution, but higher resolution is needed.

NGC~1614 has been observed with  Very Long Baseline Interferometry (VLBI) by
Corbett \etal~(2002), who did not find any unresolved radio core. Hill \etal~(2001) used long-baseline interferometry with The Australia Telescope
and reported that less than 0.9\% of the flux density at 2.3 GHz is
associated with an unresolved radio core.

 In Fig.~\ref{merlinandvla} the
1.4 GHz continuum is more centrally peaked as compared to the 5 GHz
continuum, which instead shows a ring with a radius of $\sim
1''$. We interpret this as that the 1.4 GHz continuum is mostly related
to supernova remnants, and traces old star formation, while the 5 GHz
and 8.4 GHz
continuum is more related to free-free emission from HII-regions, and
thereby traces current star formation. The ages of the starbursts can
be estimated using the spectral index, $\alpha$, (assuming S$\propto\nu^{\alpha}$) as calculated between the 1.4
GHz 
and 5 GHz maps (the upper panels in Fig.~\ref{merlinandvla}) toward the 
relevant areas  (e.g. Hirashita \& Hunt, 2006). Toward the center  the spectral index is $\alpha \sim
-0.9$, which indicates clearly synchrotron dominated emission and an age
of the starburst exceeding $\sim$ 10 Myr. Although the fidelity of the
MERLIN 1.4 GHz map is poor, the spectral index in the star forming ring
is significantly different and is  $\alpha \sim0$, which indicates thermal
emission and a starburst age of $\lapprox $5 Myr.

We conclude that our radio continuum observations are in
good agreement with the notion of an old nuclear starburst, and younger
starbursts in a  $R = 300$ pc ring.

\subsection{Is there an AGN?}

We see no direct evidence of an AGN in our data. However, the presence of an AGN can not be
ruled out, and is not in direct contradiction with our observations. The middle peak in
the MERLIN 1.4 GHz (and in the MERLIN + VLA 5 GHz) radio continuum appears close to the center of the star
forming ring, and could be associated with an AGN. Observations at
higher angular resolution and sensitivity are needed to resolve this
issue, although neither  Hill \etal (2001) nor Corbett \etal~(2002)
found any clear indications of an AGN (see sect.~\ref{nuclearradio}).
 
Risaliti \etal~(2000) report on an X-ray nucleus obscured by a
Compton thick column of gas ($N_H > 10^{24}$~$\rm{cm^{-2}}$), based on
the power law of the X-ray continuum. However, low signal to noise
measurement of a power law continuum is difficult to interpret, and
X-ray binaries have power law energy distributions that  can mimic an
AGN (e.g. Ebisawa \etal, 2008).
Wilson \etal~(2008) have observed
NGC~1614 with the submillimeter array (SMA) and suggests a nuclear non
thermal component based on ratio of submillimeter fluxes. However, the
non thermal component could be associated with either an AGN or with SNRs.

The central peak in our  CO position-velocity diagram
(Fig.~\ref{comomsandpv}, right panel) could
be associated with a nuclear, Compton thick region, obscuring an
AGN. We have calculated the largest possible radius of a Compton
thick region  ($N_H > 10^{24}$ $\rm{cm^{-2}}$) with a spherical
geometry. We assumed that at most 30\% of the flux (corresponding to $4.0 \times 10^8 $
M$_{\odot}$) in the position-velocity diagram (the central peak) could
be associated with such a region. The largest possible radius, with
$N_H > 10^{24}$ $\rm{cm^{-2}}$ would then be 140 pc ($0.''5$), and it would not be resolved in in
our CO data. Higher angular resolution would be needed to resolve the
Compton thick region. However, based on the lack of obvious signs of an AGN,
we conclude that the LINER-like optical spectrum of NGC~1614 is
probably related
to nuclear starburst activity, and not to an AGN. 

Alonso-Herrero \etal~(2000) has described a scenario where a  LINER-like optical spectrum would
be the consequence of the shock-heating by supernovae of the surrounding gas after a burst of star
formation. This would typically occur 10--15 Myr after the onset of the starburst. NGC~1614 is likely
an example of this scenario, where the nuclear (45 pc) starburst has been reported to have an age $>$ 10  Myr
(AH2001). The outflow, or super wind, described in Sect.~\ref{h1discussion} may also be related to the shock fronts giving
rise to the LINER-like spectrum. A similar scenario was suggested to
cause the LINER-like spectrum in the LINER galaxy NGC~5218 (Olsson \etal, 2007).

\subsection{Feeding of the central region}

In the position velocity diagram of the CO-data (Fig.~\ref{comomsandpv},
right panel), three
peaks are resolved. The southern and northern peaks (associated with the
star forming ring) are stronger than
the central peak. We estimate that 25\% of the gas mass in the central
beam (corresponding to $3.3 \times 10^8 $
M$_{\odot} $) is associated with the central peak. This suggests that
there has been a substantial inflow of gas to smaller radii than the star forming ring at
$R=1''$ in NGC~1614. 

Based on the large scale optical and H{\sc i} morphology,
with a tidal tail and plumes, we speculate that the tidal interaction
has triggered the formation of a bar where the gas has been transported
along the leading edge and piled up in the star forming ring. An inner
bar has decoupled dynamically at this radius, and allowed the gas to
flow toward the center. Star formation has
occurred in this dynamically decoupled bar, and can be seen in the 1.4
GHz MERLIN map (Fig.~\ref{merlinandvla}). Inflow occuring before the formation
of the ring, and the continued inflow along a decoupled bar after the ring formation,
has triggered and maintained the nuclear activity of NGC~1614.

\subsection{The nature of the crossing dust lane and the interaction}
\label{dust_lane}

Prominent dust lanes in merging galaxies are not uncommon and have
a variety of properties. One example is the dust lane of the NGC~4194 merger
(e.g. Aalto and H\"uttemeister 2000).
Just like the NGC~1614 dust lane, it is a minor axis dust lane crossing the
galaxy at an angle to the velocity field. 

In the NGC~4194 merger it
is speculated that the dust lane is feeding the central region with H$_2$ for
star formation from the outer regions of the merger. A similar scenario is possible
for NGC~1614 and should be studied at higher resolution. It is interesting
to speculate whether the minor axis dust lane is part of an inner polar ring
structure.

In a numerical simulation study of minor- or intermediate mergers, Bournaud \etal\ (2005) found that
the gas brought in by the disturbing companion galaxy is generally found at large radii in the merger
remnant. The gas is returning to the system from tidal tails and often forms rings -- polar, inclined
or equatorial -- that will appear as dust lanes when seen edge-on.

The CO emission we identify as associated with the fore-ground dust lane is blueshifted with respect to the
expected rotational velocity at the same position in the plane of the galaxy. This is consistent with it
not being in the same plane, but does not constitute evidence. Alternatively the dust lane acts as a gaseous bar, inclined to the 
K-band stellar bar, along which gas is being funneled to the center. This interpretation is not consistent with
the CO velocity field however. Higher resolution molecular observations are necessary to distinguish between the 
polar ring/gas bar scenario and to investigate how the gas is flowing towards the nucleus.

Both a polar ring and a bar can be triggered by the interaction just like the tidal features of NGC~1614.
Neff \etal\ (1990) suggest that an intruding companion is located 10$''$ to the south-west of the center
of NGC~1614. Peculiar H$\alpha$ velocities and an alignment of the jet-like optical tail suggests this.
An inspection of the HST WFPC2 F606W image does indeed reveal a drawn-out feature here pointing in the direction of
the tail. We find no molecular counterpart to this object, but this is hardly surprising if it has already been
stripped of its gas content and only has its nucleus left. The feature also appears in the K-band image.
AH2001 suggest that there is evidence of the nucleus of a companion approximately 1$''$ north-east of the nucleus
in their B-band image. There is no hint of this companion in the K-band image - and no evidence of a particular disturbance
in our CO or HI velocity field here, although the spatial resolution needs to be higher to be certain.

\section{Conclusions}

\begin{enumerate}

\item We detect  $ 3 \times 10^9  \, \rm
      M_{\odot}$ of molecular gas within a north to south slightly elongated
      bar like structure with a size of $\sim 2.2 \times 1.5$ kpc.  

\item The north to south rotation of the central kpc molecular gas is consistent with an inclined disc /
      bar that follows the expected rotational direction based on the
      assumption of trailing spiral arms.

\item There is a molecular extension to the northeast,
       associated with a crossing dust lane.

\item The molecular gas in the central kpc is double peaked, with peaks
      at $R=300$ pc. There is a third, central peak at lower integrated
      intensities, probably associated with a nuclear starburst/AGN.

\item We have detected a  MERLIN and VLA 5 GHz radio continuum
      ring with a radius of 300 pc. This ring is coincident with previous
      radio continuum and Pa$\alpha$
      observations. We conclude  that the radio continuum ring originates in recent
      star formation. The CO peak intensity is shifted 1$''$ to the north-west of the 
      center of the starburst ring.

\item The MERLIN 1.4 GHz radio continuum is triple peaked, with a peak
      separation of $\sim 1''$ (300 pc). The brightest peaks occur in
      the northern and southern part of the
      star forming ring with a radius of 300 pc. The other peak, which is almost as bright, occur
      inside the star forming ring, close to its center. The brightness temperature of maximum
      53 000 K of the peaks are consistent with a cluster of supernova
      remnants, although an AGN can not be ruled out for the peak close to the center.

\item The position velocity diagram of the  H{\sc i}-absorption was
      used to calculate a dynamical mass of  $ 3.5 \times 10^9  \, \rm
      M_{\odot}$ in the central $R=1''$ (300 pc). An upper limit to the molecular gas mass
      in the same region is $\sim 1.3 \times 10^9  \, \rm
      M_{\odot}$. This gives a gas mass fraction of $\sim 0.4$ or less in the
      central $R=1''$ (300 pc).

\item A solid body rotation curve was fitted to the H{\sc i}-absorption
      velocity field, out to a radius of $2''$ (600 pc). The residuals are blueshifted over all of the
      fitted area. We interpret this as an outflow or a superwind, with
      a deprojected velocity of $\sim$160 \kms.

\item  We suggest that gas may flow to the nucleus along a dynamically decoupled
bar inside the star forming ring (or spiral), which we see tentative evidence for in the 1.4 GHz MERLIN map.
The existence of such a bar should be investigated with CO observations and higher angular resolution.

\item The LINER like spectrum is likely related to the shock fronts
      associated with either the large scale super wind, or with shock
      fronts on smaller scales related to supernova remnants from
      previous bursts of star formation. We conclude that the LINER activity
      observed in NGC~1614 is probably due to starburst activity, and not to AGN-activity. 

\end{enumerate}

\begin{acknowledgements} 

We gratefully acknowledge use of the NRAO Very Large Array (VLA), the Owens Valley Radio Observatory 
(OVRO), the Multi-Element Radio Linked Interferometer Network (MERLIN),
their associated data archives and the Hubble Space telescope ({\it {HST}})
data archive. We thank Waykin Ariyasoonthorn for his help with the
OVRO CO data reduction. We thank Rainer Beck and Alessandro Romeo for
useful discussions regarding the interpretation of the radio continuum results. 
This research has made use of the NASA/IPAC Extragalactic Database
(NED) which is operated by the Jet Propulsion Laboratory, California
Institute of Technology, under contract with the National Aeronautics and
Space Administration. Parts of this work was supported by the EU Marie Curie Training Site programme
under contract no. HPMT-CT-2000-00069 (JTRA).

\end{acknowledgements}

\end{document}